\documentclass[prl,twocolumn,showpacs,amsmath,amssymb,superscriptaddress]{revtex4}
\usepackage{graphicx}
\usepackage{dcolumn}
\usepackage{bm}
\usepackage{epsfig}
\begin{document}


\title{
Controlling magnetic order and quantum disorder in one- and zero-dimensional molecule-based magnets}

\author{T. Lancaster}
\affiliation{Durham University, Centre for Materials Physics, South Road, 
Durham, DH1 3LE, United Kingdom}
\author{P.A. Goddard}
\affiliation{University of Warwick, Department of Physics, Coventry,
  CV4 7AL, UK}
\author{S.J. Blundell}
\author{F.R. Foronda} 
\author{S. Ghannadzadeh}
\author{J.S. M\"{o}ller}
\affiliation{Oxford University Department of Physics, Clarendon Laboratory, 
Parks Road, Oxford, OX1 3PU, United Kingdom}
\author{P.J. Baker}
\author{F.L. Pratt}
\affiliation{ISIS Facility, Rutherford Appleton Laboratory, Chilton, 
Oxfordshire OX11 0QX, United Kingdom}
\author{C. Baines}
\affiliation{Laboratory for Muon-Spin Spectroscopy, Paul Scherrer Institut, CH-5232 Villigen PSI,  Switzerland }
\author{L. Huang}
\author{J. Wosnitza}
\affiliation{Dresden High Magnetic Field Laboratory, Helmholtz-Zentrum Dresden-Rossendorf, D-01314 Dresden, Germany}
\author{R.D. McDonald}
\author{K.A. Modic}
\author{J. Singleton}
\author{C.V. Topping}
\affiliation{National High Magnetic Field Laboratory, Los Alamos National Laboratory, MS-E536, Los Alamos, New Mexico 87545, USA}
\author{T.A.W. Beale}
\author{F. Xiao}
\affiliation{Durham University, Centre for Materials Physics, South Road, 
Durham, DH1 3LE, United Kingdom}
\author{J.A. Schlueter}
\affiliation{Materials Science Division, Argonne National Laboratory, Argonne, Illinois 60439, USA}
\author{R.D. Cabrera}
\author{K.E. Carreiro} 
\author{H.E. Tran}
\author{J.L. Manson}
\affiliation{Department of Chemistry and Biochemistry, Eastern Washington
University, Cheney, Washington 99004, USA}

\date{\today}

\begin{abstract}
We investigate the structural and magnetic properties of two
molecule-based magnets synthesized  from the same starting
components. 
Their different structural motifs promote contrasting exchange pathways and consequently lead to
markedly different magnetic ground states. Through examination of their structural
and magnetic properties we show that [Cu(pyz)(H$_{2}$O)(gly)$_{2}$](ClO$_{4}$)$_{2}$ may be considered
a quasi-one-dimensional quantum Heisenberg antiferromagnet while the related compound [Cu(pyz)(gly)](ClO$_{4}$),
which is formed from dimers of antiferromagnetically interacting
Cu$^{2+}$ spins, 
remains disordered down to at least 0.03~K in zero field, but shows
a field--temperature phase diagram reminiscent of that seen in
materials showing a Bose--Einstein condensation of magnons. 
\end{abstract}
\pacs{75.30.Et, 74.62.Bf, 75.50.Ee, 75.50.Xx}
\maketitle

A goal of research in the field of molecular magnetism is to gain control of chemical components in order that
desirable magnetic behavior may be achieved  \cite{batten, paul}. This relies on a detailed understanding of the relationship between
starting materials, structure and magnetic properties. Here we present a case where a synthetic route leads to the
realization of two structurally distinct materials based on similar
chemical components but different low-dimensional motifs. 
Through magnetometry, heat capacity, electron paramagnetic resonance (EPR) and muon-spin relaxation ($\mu^{+}$SR) we show that these materials represent realizations of
different models of quantum magnetism in reduced dimensions, namely
(i) the quasi one-dimensional $S=1/2$  quantum 
Heisenberg antiferromagnet (1DQHAF)  \cite{giamarchi} and 
(ii) an assembly of  weakly coupled singlet dimers (each one composed
of two antiferromagnetically coupled $S=1/2$ spins) with a quantum disordered ground state
which may be driven with an applied magnetic field via a quantum
critical point (QCP) into a magnetic phase reminiscent of the Bose--Einstein
condensation (BEC) of magnons  \cite{ruegg}. 

The synthesis of the materials (described in full in the supplemental information) involves 
mixing aqueous solutions of Cu(ClO$_{4}$)$_{2} \cdot \ 6$H$_{2}$O,
glycine [$\equiv$NH$_{2}$CH$_{2}$COOH, (gly)], and pyrazine [$\equiv$C$_{4}$H$_{4}$N$_{2}$, (pyz)]. 
Purple blocks of [Cu(pyz)(gly)](ClO$_{4}$) typically form first upon 
slow evaporation of the solvent, while continued 
evaporation leads to the formation of blue rods of [Cu(pyz)(H$_{2}$O)(gly)$_{2}$](ClO$_{4}$)$_{2}$.
Despite the presence of the same molecular components, the infrared spectra of these materials differ markedly 
between 1300 and 1700~cm$^{-1}$ allowing their identification and isolation.  Varying the relative ratios of chemical reagents 
does not alter the outcome and the latter material is always obtained in higher yield. 

\begin{figure}
\begin{center}
\epsfig{file=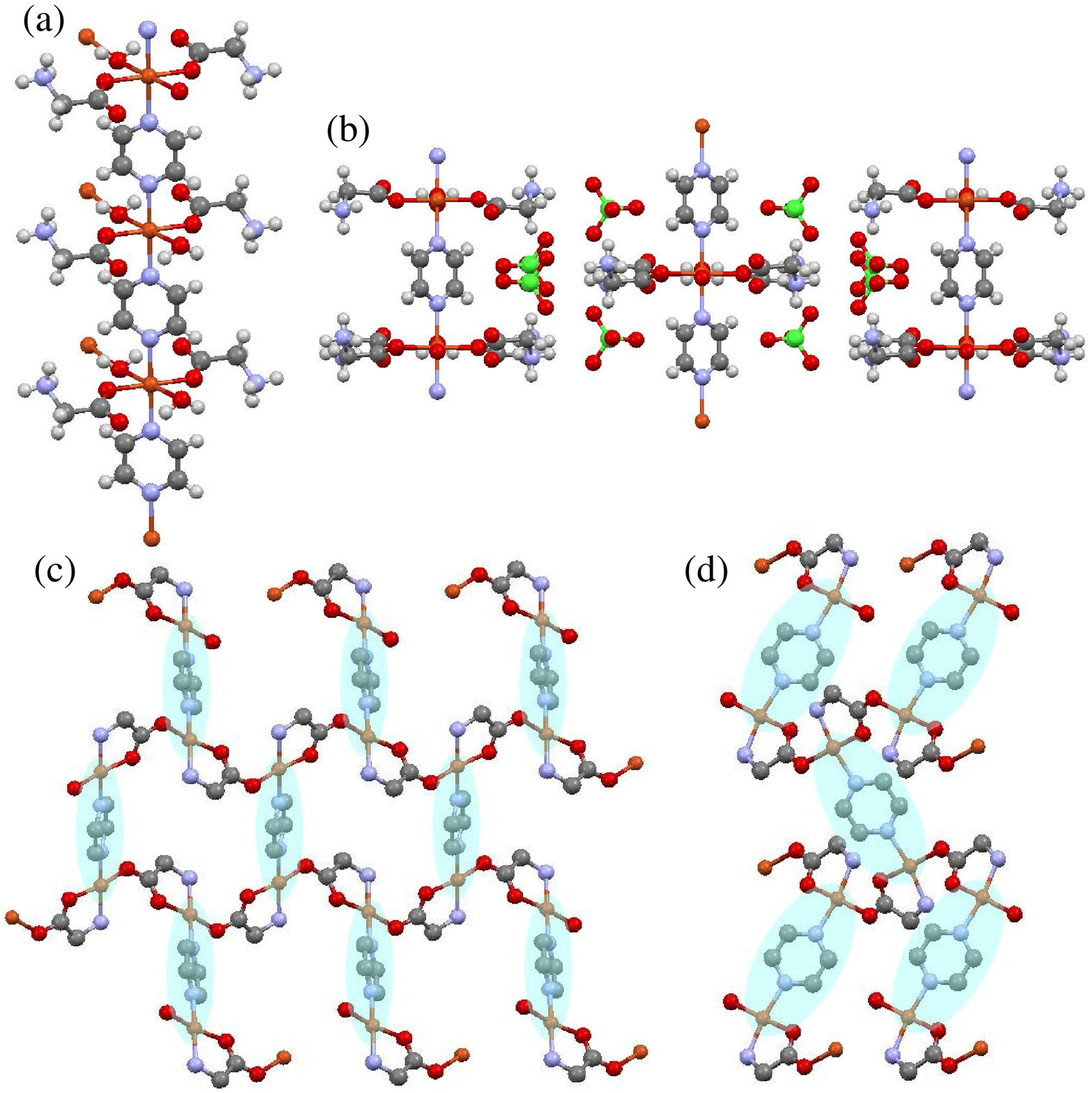,width=\columnwidth}
\caption{
(a) Cu-(pyz)-Cu chains on which
[Cu(pyz)(H$_{2}$O)(gly)$_{2}$](ClO$_{4}$)$_{2}$ is based. Water and
gly groups which coordinate with the Cu ions are also shown. (b) The material viewed
along the $c$-axis showing the packing of the chains with
non-coordinating ClO$_{4}$ groups. 
 [Cu(pyz)(gly)](ClO$_{4}$) viewed along (c) the  $a$-axis and (d) the
 $c$-axis showing pairs of Cu$^{2+}$ ions strongly coupled through
pyz ligands to form alternating dimers. (Dimers are shaded and the 
ClO$_{4}$ ions have been removed for clarity.)
\label{struc}}
\end{center}
\end{figure}

The structure of [Cu(pyz)(H$_{2}$O)(gly)$_{2}$](ClO$_{4}$)$_{2}$,
which crystallizes in space group $C2/c$, is
based on linear chains of $S=1/2$ Cu$^{2+}$ ions
linked with pyz ligands as shown in Fig.~\ref{struc}(a).  Glycine groups and H$_{2}$O molecules
coordinate with the Cu ions and these, along with non-coordinating
ClO$_{4}$ counter-ions, 
act to separate the chains [Fig.~\ref{struc}(b)]. The pyz ligand is known to be an effective
mediator of magnetic exchange in materials of this type and so we
would expect the chain-like structure to promote one-dimensional (1D)
antiferromagnetic behavior. 
In contrast, the structure of [Cu(pyz)(gly)](ClO$_{4}$) (space group $P2_{1}/n$) is based on a
lattice of 
alternating Cu$^{2+}$ dimers as shown in Fig.~\ref{struc}(c) and (d). 
The two Cu$^{2+}$ ions in a dimer are coupled
by a pyz ligand and these dimers are tethered 
with gly bridges that connect the dimers to form corrugated sheets,
with non-coordinating ClO$^{-}_{4}$ ions lying between these sheets. The exchange through gly
groups and ClO$^{-}_{4}$ ions might be expected to be comparatively weak,
suggesting that the physics of this material is due to dimer units
weakly coupled with their neighbors.

\begin{figure}
\begin{center}
\epsfig{file=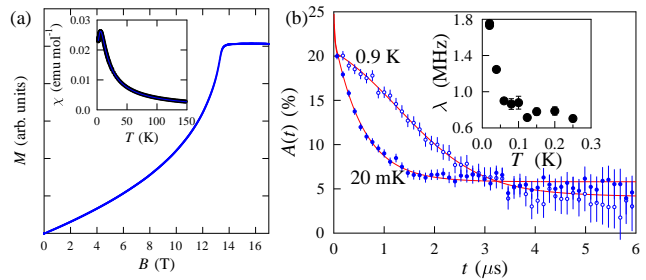,width=\columnwidth}
\caption{
(a) Magnetization of chain-like [Cu(pyz)(H$_{2}$O)(gly)$_{2}$](ClO$_{4}$)$_{2}$ measured at 0.5~K. {\it Inset:}
magnetic susceptibility measured in an applied field of 100~mT. 
(b) ZF $\mu^{+}$SR spectra measured at 0.9 and 0.02~K. {\it Inset:}
Relaxation rate 
as a function of temperature showing the magnetic transition. 
\label{data}}
\end{center}
\end{figure}

The magnetic behavior of chain-like
[Cu(pyz)(H$_{2}$O)(gly)$_{2}$](ClO$_{4}$)$_{2}$ was 
characterized using magnetic susceptibility and 
magnetization measurements on polycrystalline samples 
and found to be well described by the
predictions of the 1DQHAF model.
This model is
defined by a Hamiltonian
\begin{equation}
H = J \sum_{\langle i,j \rangle \parallel}
\boldsymbol{S}_{i}\cdot\boldsymbol{S}_{j}+
J_{\perp} \sum_{\langle i,j \rangle \perp}
\boldsymbol{S}_{i}\cdot\boldsymbol{S}_{j} - g \mu_{\mathrm{B}} B
 \sum_{i}S^{z}_{i},
\label{1daf}
\end{equation}
where $J$ is the strength of the
exchange coupling within the magnetic chains, $J_{\perp}$
is the coupling between chains, and the first and
second summations refer to summing over unique pairs of
nearest neighbors parallel and perpendicular to the chain, respectively.
The magnetic susceptibility [shown inset in
Fig.~\ref{data}(a)] is well described by the form expected  \cite{johnston}
for the model in Eq.~(\ref{1daf})
with antiferromagnetic (AF) intrachain exchange strength
$|J|=9.4(1)$~K and $g=2.10(1)$. 
Magnetization measurements are shown in
Fig.~\ref{data}(a) and have a concave curvature typical of a quasi-1D
system. The saturation field of the magnetization,
$B_{\mathrm{c}}=13.3(1)$~T,  allows us to estimate
the exchange since we expect $g \mu_{\mathrm{B}} B_{\mathrm{c}} =
2J+4J_{\perp}$. 
Assuming $J_{\perp}/J \ll 1$, suggests $J= 9.4(1)$~K in agreement with
the results above. We note that this value is quite typical of exchange
through Cu-pyz-Cu bonds  \cite{paul,goddard_njp}. 

%

In order to assess how well
[Cu(pyz)(H$_{2}$O)(gly)$_{2}$](ClO$_{4}$)$_{2}$
approximates a 1DQHAF we use the ratio of the magnetic ordering
temperature $T_{\mathrm{N}}$ to $J$. This should be zero for an ideal
1D magnet and close to unity for an isotropic system. 
Many thermodynamic probes cannot resolve the ordering
transition in 1D systems
owing to strong thermal and quantum fluctuations, which lower the magnitude of $T_{\mathrm{N}}$, 
resulting in correlations with sizeable correlation length building up in the chains just above the ordering temperature
and also reduce the magnitude of the magnetic moment  \cite{sengupta}. We
have shown previously  \cite{tom,paul} 
that muons are often
sensitive to LRO that can be very difficult to detect with
thermodynamic probes in quasi 1D systems. 

Example zero-field (ZF) $\mu^{+}$SR spectra are shown in Fig.~\ref{data}(b). 
Although oscillations, characteristic of a quasi-static local magnetic field at the 
muon stopping site, are not observed at any temperature, we do see a
sudden change in behavior at low temperature that is evidence for
magnetic order. 
Data were well described across the measured temperature range by a function
\begin{equation}
A(t) = A_{1}e^{-\sigma^{2}_{1} t^{2}}  + A_{2}e^{-\sigma_{2}^{2} t^{2} }+
A_{3}e^{-\lambda t}+ A_{\parallel},
\end{equation}
where the first term captures the rapid relaxation observed at early times and
the term with amplitude $A_{2}$ captures the weak relaxation due to disordered nuclear
moments. As the temperature is lowered
we observe a sharp increase in both the relaxation rate $\lambda$
(reflecting the behavior of electronic
moments) and
the baseline amplitude $A_{\parallel}$ around 40~mK. This behavior, which has been observed
previously
in chain-like materials of this type  \cite{paul}, is strongly indicative of
magnetic order. From this we estimate $T_{\mathrm{N}} = 0.04(1)$~K for
[Cu(pyz)(H$_{2}$O)(gly)$_{2}$](ClO$_{4}$)$_{2}$. 

Quantum Monte Carlo simulations provide a means of estimating
the effective interchain coupling $J_{\perp}$ in a quasi-1D antiferromagnet via the 
expression  \cite{yasuda}
\begin{equation}
|J_{\perp}| = \frac{T_{\mathrm{N}}}{ 4c \sqrt{ \ln\left( \frac{a J}{T_{\mathrm{N}}}\right)
+ \frac{1}{2}\ln \ln \left(  \frac{a J }{T_{\mathrm{N}}} \right)  }  },
\label{yasj}
\end{equation}
where $c = 0.233$ and $a = 2.6$ for $S=1/2$ spins.  Using our values of
$T_{\mathrm{N}}$ and $J$
we obtain an estimate $|J_{\perp}/J|\approx 2\times 10^{-3}$
which provides a measure of the degree to which
[Cu(pyz)(H$_{2}$O)(gly)$_{2}$](ClO$_{4}$)$_{2}$ 
realizes the 
1DQHAF (for which $|J_{\perp}/J|=0$). It  therefore provides a more isolated
realization of the
1DQHAF than Cu(pyz)(NO$_{3}$)$_{2}$ ($|J_{\perp}/J|=4.4\times
10^{-3}$)  \cite{tom}, 
but is less well isolated than
Sr$_{2}$CuO$_{3}$ ($|J_{\perp}/J|=7\times 10^{-4}$)  \cite{keren} 
and DEOCC-TCNQF$_{4}$ ($|J_{\perp}/J| < 6\times 10^{-5}$)  \cite{francis}.

\begin{figure}
\begin{center}
\epsfig{file=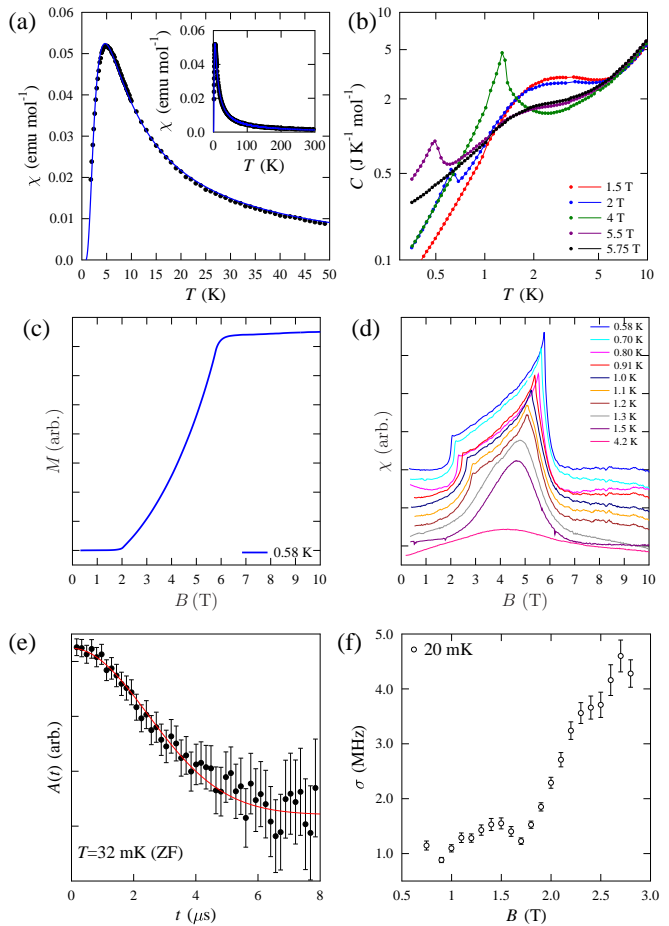,width=\columnwidth}
\caption{
(a) Static magnetic susceptibility data for [Cu(pyz)(gly)](ClO$_{4}$) measured
in a field of 5~mT. A fit
is shown to the Bleaney-Bowers model of non-interacting dimers. 
(b) Heat capacity showing magnetic transitions in applied magnetic fields.
(c) Magnetization at 0.58~K and (d) dynamic susceptibility at several
temperatures
showing sharp features at the phase transitions (data offset for each
temperature for clarity).
(e) Example ZF $\mu^{+}$SR spectrum 
measured at 32 ~mK showing no signature of long-range magnetic order. 
(f) Relaxation rate at 20~mK
as a function of transverse applied magnetic  field showing a transition around
1.7~T. 
\label{data3}}
\end{center}
\end{figure}

We now turn to the magnetic properties of  [Cu(pyz)(gly)](ClO$_{4}$). The results of magnetic susceptibility
measurements on single crystal samples are shown in
Fig.~\ref{data3}(a) with a magnetic field applied 
along the long axis of a crystallite. 
These data are well
described by the Bleaney-Bowers model  \cite{bleaney} which gives the susceptibility of a 
system of isolated antiferromagnetically coupled dimers and yields an intradimer exchange strength of
$|J_{0}|=7.5(1)$~K. 
 A slightly better fit may be
obtained assuming a mean-field ferromagnetic (FM) interdimer coupling,
resulting in AF intradimer coupling of $|J_{0}|=8.1(1)$~K and
FM coupling $J\approx 2$~K. We note that such fits are quite sensitive to the details of
the model used (we return to the nature of this coupling below). 
Our ZF $\mu^{+}$SR measurements made down to 32~mK
[Fig.~\ref{data3}(e)] show no indication
of magnetic order, or sizeable relaxation due to fluctuating
electronic moments
 with the spectra remaining typical of relaxation
due to disordered nuclear magnetism.

These results suggest that [Cu(pyz)(gly)](ClO$_{4}$)
should be described via a Hamiltonian
\begin{equation}
H =  J_{0}\sum_{i} \boldsymbol{S}_{1,i}\cdot \boldsymbol{S}_{2,i} +
\sum_{\langle  mnij \rangle} J_{mnij} \boldsymbol{S}_{m,i}\cdot
\boldsymbol{S}_{n,j}
- g \mu_{\mathrm{B}} B \sum_{\langle ni \rangle} S^{z}_{m,i}, 
\end{equation}
where $i,j$ label dimers and $m,n=1, 2$ label their magnetic sites
 \cite{ruegg}.
 If $J_{mnij}$ are weak compared to the antiferromagnetic
intradimer exchange $J_{0}$ then this 
causes the ground state to be one of quantum disorder, 
formed from an array of $S=0$ spin singlets. 

The application of  a magnetic field is found to drive this material
through a magnetic phase transition. 
The transition is seen in heat capacity, dynamic
magnetic susceptibility
 and $\mu^{+}$SR measurement made on [Cu(pyz)(gly)](ClO$_{4}$).
 As shown in
Fig.~\ref{data3}(b) sharp peaks are observed in heat capacity
measurements on a single crystal (with the
field applied as for the susceptibility) in the $2$--$6$~T
region in scans across the  temperature range $0.4 < T < 1.4$~K. 
Single crystal, dynamic magnetic susceptibility measurements 
were
performed using a radio-frequency based susceptometer
 \cite{ghannadzadeh, suppl}.
The dynamic  susceptibility, $\chi = {\rm
  d}M/{\rm d}B$,  was measured in two different orientations, with the
magnetic field applied close to normal to the (110) or (122) crystallographic planes. 
The field for the (110) orientation is parallel to the $a$-$b$ plane, which contains the dimers.
The phase
transitions may be identified from the sharp features in
 $\chi$, shown in Fig.~\ref{data3}(d).
The magnetization at 0.58 K, found from
integration of $\chi$, is shown in Fig.~\ref{data3}(c). 
Transitions are also observed in $\mu^{+}$SR (for which an unaligned
polycrystalline sample was measured) using both a transverse
field (TF)
geometry (with initial muon spin perpendicular to the applied field)
 and in  a
longitudinal field (LF) geometry 
(initial muon spin parallel to the applied field) 
 \cite{suppl}. The form of the transition in the
TF geometry  [Fig.~\ref{data3}(f)] involves a rapid rise in relaxation and 
is similar to that observed for field-induced transitions in molecular
magnets
of this sort  \cite{steele} and also in the candidate BEC material
Pb$_{2}$V$_{3}$O$_{9}$  \cite{carlo}. 
In the LF data the phase boundary is identified via a sudden change in the integrated asymmetry
and in the relaxation rate \cite{suppl}. 

The positions of the phase boundaries determined by the different
measurements show some degree of dependence on the crystal orientation, reflecting
the effect of $g$-factor anisotropy. 
Our EPR measurements  \cite{suppl} allow us to determine the $g$-factor for
fields applied normal to the (110) plane as $g^{(110)}=2.18$.
 The phase boundaries are found to
coincide if we take $g^{(122)}=2.15$
for the dynamic susceptibility with the other measured crystal orientaton,
$g^{\mu\mathrm{SR}}=2.20$ for $\mu^{+}$SR and
$g^{\mathrm{HC}}=2.30$ for heat capacity. (We note that these $g$-factors all fall
withing the range typically found in Cu-based coordination polymers \cite{paul,goddard_njp}.)
Scaling the field $B_{\mathrm{c}}$ at which the phase change occurs
for each measurement by plotting $g \mu_{\mathrm{B}}B_{\mathrm{c}}$,
we obtain the phase diagram shown in Fig.~\ref{data2}. 
which, as shown below, is consistent with that of a system of AF
dimers, with weak FM interdimer coupling  \cite{tachiki}. 

The phase diagram results from the fact that at temperatures well below the 
intradimer separation $|J_{0}|$, the ground state is a
quantum-disordered paramagnet formed from a sea of singlets  \cite{tachiki,ruegg,zheludev,tsvelik}. The
applied field (which we assume is along $z$) closes the singlet-triplet spin gap at a
QCP at $g \mu_{\mathrm{B}} B_{\mathrm{c}1}$, leading to
a state of LRO formed from the transverse spin components $\langle
S_{x}\rangle$ and $\langle S_{y}\rangle$, which spontaneously break the
$O(2)$ symmetry of the spin Hamiltonian. Further application of the
field for $B>B_{\mathrm{c1}}$ cants the spin components along $z$ until we encounter the fully
$z$-polarized FM phase beyond another QCP at $g\mu_{\mathrm{B}}B_{\mathrm{c}2}$. 
  In the mean-field
approximation, the upper phase boundary for FM coupled dimers
occurs at the intradimer exchange value $g \mu_{\mathrm{B}}B_{\mathrm{c2}}=|J_{0}|$ which we estimate
from the phase boundary to be $9.0(2)$~K, broadly consistent with, but
slightly larger than, the
value derived from dc susceptibility, but in agreement with the expected value for
AF exchange mediated by a pyz group (such as that in
[Cu(pyz)(H$_{2}$O)(gly)$_{2}$](ClO$_{4}$)$_{2}$ above). The mean-field model also
predicts that $g \mu_{\mathrm{B}}B_{\mathrm{c}1} = |J_{0}| - z|J_{1}|/2$,
where $z$ is the number of interacting nearest neighbors linked with
a mean-field FM interdimer exchange constant $J_{1}(= \langle J_{mnij}\rangle)$ . 
We find $g \mu_{\mathrm{B}}B_{\mathrm{c}1}=2.5(1)$~K and so, assuming $z=4$, we obtain
$|J_{1}|\approx 3.3(1)$~K. If, instead, an AF interdimer coupling is
assumed  \cite{tachiki}
we would obtain $|J_{0}|=4.5$~K, which is not compatible with the
dc susceptibility measurement. 

\begin{figure}
\begin{center}
\epsfig{file=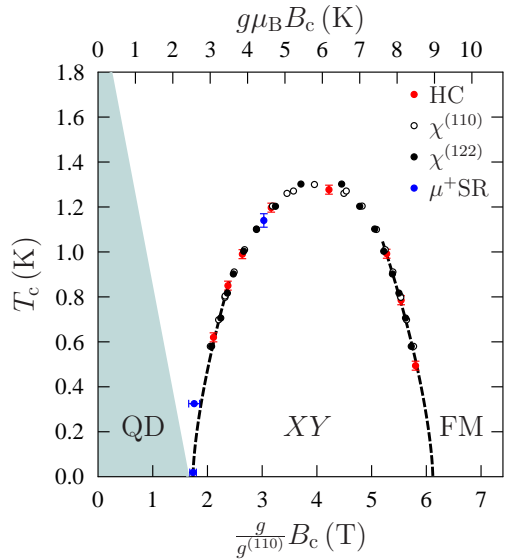,width=7cm}
\caption{
The $B$--$T$ phase diagram for [Cu(pyz)(gly)](ClO$_{4}$), showing quantum
critical points at $g\mu_{\mathrm{B}}B_{\mathrm{c}}=2.5(1)$~K and
9.0(2)~K.
Dashed lines show the expected behavior of the phase boundaries with
exponent
$\phi=2/3$, predicted for the BEC of magnons model. 
\label{data2}}
\end{center}
\end{figure}

Dimer systems such as these are often discussed in the context of Bose--Einstein
condensation (BEC) of magnons, which provides a similar description of the
physics in terms of a transition from a triplon vacuum at low field to
a Bose condensed state 
that breaks a global $U(1)$ symmetry, 
 isomorphic to the $O(2)$ invariance of the isolated dimer system in
 magnetic field
  \cite{ruegg,zheludev,tsvelik}. 
The BEC picture predicts phase boundaries with a $\phi=2/3$ power law
exponent, which is not inconsistent with our data, although the
paucity of data points in the critical region prevents this from being
rigorously assessed. 
Although this material may provide a further approximate realization of
magnon BEC, we note that $O(2)$ [ $\equiv U(1)$] symmetry is essential for the
realization of the model and
we cannot rule out a small Dzyaloshinsky-Moriya (DM) interaction  \cite{yoshida} 
providing
a term in the Hamiltonian $\boldsymbol{D}\cdot(\boldsymbol{S}_{1}\times \boldsymbol{S}_{2})$  
that may break the $XY$ symmetry in the
$a$-$b$ plane. Although
inversion centers exist between neighboring dimer centers with the same
orientation shown in Fig.~\ref{struc}(d), the neighbors with
different orientations are symmetry related through a glide plane. 
This implies a non-zero DM interaction with strength $\boldsymbol{D}$ within the 
$a$-$c$ plane.

In conclusion, we have shown how a chemical synthesis route leads to
two separable phases of matter whose difference in structure allows
them to realize two distinct models of quantum magnetism distinguished
by their dimensionality. The first is
a good realization of a $S=1/2$ 1DQHAF which shows magnetic order
at very low temperature. The other is based on magnetic dimers giving
rise to a quantum disordered ground state and field induced
$XY$ antiferromagnetic phase. This work demonstrates the potential for
creating still more exotic magnetic ground states from coordination
polymers such as Cu-pyz systems. Indeed, through chemically engineering
frustration into such a system, a spin-liquid ground state might be
achieved. 

Part of this work was carried out at the STFC ISIS facility and at the Swiss Muon Source, 
Paul Scherrer Institute, Switzerland and we are grateful for the provision of beamtime. 
This work is supported by EPSRC (UK). We are grateful to R. Williams
and D.S Yufit for
experimental assistance. 
Work at EWU was supported in part by the U.S. National Science Foundation under grant no. DMR-1005825.
The NHMFL is supported by NSF, DoE and the State of Florida.

\end{document}



\title{Supplemental information}

\author{T. Lancaster}
\affiliation{Durham University, Centre for Materials Physics, South Road, 
Durham, DH1 3LE, United Kingdom}
\author{P.A. Goddard}
\affiliation{University of Warwick, Department of Physics, Coventry,
  CV4 7AL, UK}
\author{S.J. Blundell}
\author{F.R.  Foronda} 
\author{S. Ghannadzadeh}
\author{J.S. M\"{o}ller}
\affiliation{Oxford University Department of Physics, Clarendon Laboratory, 
Parks Road, Oxford, OX1 3PU, United Kingdom}
\author{P.J. Baker}
\author{F.L. Pratt}
\affiliation{ISIS Facility, Rutherford Appleton Laboratory, Chilton, 
Oxfordshire OX11 0QX, United Kingdom}
\author{C. Baines}
\affiliation{Laboratory for Muon-Spin Spectroscopy, Paul Scherrer Institut, CH-5232 Villigen PSI,  Switzerland }
\author{L. Huang}
\author{J. Wosnitza}
\affiliation{Dresden High Magnetic Field Laboratory, Helmholtz-Zentrum Dresden-Rossendorf, D-01314 Dresden, Germany}
\author{R.D. McDonald}
\author{K.A. Modic}
\author{J. Singleton}
\author{C.V. Topping}
\affiliation{National High Magnetic Field Laboratory, Los Alamos National Laboratory, MS-E536, Los Alamos, New Mexico 87545, USA}
\author{T.A.W. Beale}
\author{F. Xiao}
\affiliation{Durham University, Centre for Materials Physics, South Road, 
Durham, DH1 3LE, United Kingdom}
\author{J.A. Schlueter}
\affiliation{Materials Science Division, Argonne National Laboratory, Argonne, Illinois 60439, USA}
\author{R.D. Cabrera}
\author{K.E. Carreiro} 
\author{H.E. Tran}
\author{J.L. Manson}
\affiliation{Department of Chemistry and Biochemistry, Eastern Washington
University, Cheney, Washington 99004, USA}

\date{\today}

\pacs{75.30.Et, 74.62.Bf, 75.50.Ee, 75.50.Xx}
\maketitle

\section{Synthesis details}

The compounds Cu(gly)(pyz)(ClO$_{4}$) (hereafter {\bf 1}) and
[Cu(gly)$_{2}$(pyz)(H$_{2}$O)](ClO$_{4}$)$_{2}$ (hereafter {\bf 2}) were synthesised as
follows. 
Following a general procedure, aqueous solutions of
Cu(ClO$_{4}$)$_{2}$·6H$_{2}$O (0.5017~g, 1.0~mmol), glycine (0.1016~g, 1.0~mmol), and
pyrazine (0.1084~g, 1.0~mmol) are slowly mixed to form a deep blue
colored solution. Purple blocks of {\bf 1} typically form first upon slow
evaporation of the solvent over a period of about one week and are
removed via suction filtration (0.0516~g). Continued evaporation of
the mother liquor leads to blue rods of {\bf 2} in higher yield (0.0951~g). Despite the presence of the same molecular components, the
infrared spectra of {\bf 1} and {\bf 2} differ markedly between 1300
and 1700~cm$^{-1}$. 
On several occasions, the two phases grew simultaneously but
could be mechanically separated and their identities readily confirmed
by IR spectroscopy and X-ray diffraction. Varying the relative ratios
of chemical reagents did not alter the outcome and {\bf 2} was always
obtained in higher yield. 

\section{X-ray structure determination of compound 1 }
A suitable crystal was selected,
attached to a glass fiber and data were collected at 298(2)~K using a
Bruker/Siemens SMART APEX instrument (Mo K$\alpha$ radiation, $\lambda$ = 0.71073~\AA)
equipped with a Cryocool NeverIce low temperature device. Data were
measured using omega scans of 0.3$^{\circ}$ per frame for 5~seconds, and a
full sphere of data was collected. A total of 2400 frames were
collected with a final resolution of 0.83~\AA. The first 50 frames were
recollected at the end of data collection to monitor for decay. Cell
parameters were retrieved using SMART \cite{jlm1}  software and refined using
SAINTPlus \cite{jlm2}  on all observed reflections. Data reduction and correction
for Lorentz polarization (Lp) and decay were performed using the SAINTPlus software. The data
were rotationally twinned and deconvoluted using CELL\_NOW \cite{jlm3}  giving
a 2.8$^{\circ}$ rotation about the real axis 1.000, 0.703, 0.784 with a
twinning ratio of 0.218(4). The matrix used to relate both
orientations is 0.988 0.013 -0.031 -0.036 1.000 0.033 0.048 -0.016
1.010. Absorption corrections were applied using TWINABS \cite{jlm4}.  The
structure was solved by direct methods and refined by least squares
methods on $F^{2}$ using the SHELXTL program package \cite{jlm5}.  The structure was
solved in the space group $P2_{1}/n$ (no.\ 14) by analysis of systematic
absences. All atoms were refined anisotropically. The perchlorate
oxygen atoms were disordered and modeled in three separate locations
with occupancies of 30, 40, 30\%. Soft restraints were applied to the
Cl-O distances and thermal parameters. No decomposition was observed
during data collection. Details of the data collection and refinement
are given in Table~1. 

\begin{table*}
\begin{tabular}{cc}
\hline
Empirical formula &	C$_{4}$H$_{6}$ClCuN$_{2}$O$_{6}$ \\
Formula weight &	277.10 \\
Temperature &	298(2) K \\
Wavelength &	0.71073 \AA\\
Crystal system 	&Monoclinic\\
Space group &	$P2_{1}/n$\\
Unit cell dimensions&	$a = 7.2054(15) $~\AA	\\
&	$b = 14.282(3)$~\AA\\	
&	$c = 8.890(2)$~\AA	\\
&$\alpha$= 90$^{\circ}$\\
&$\beta$= 105.986(4)$^{\circ}$\\
&$\gamma$ = 90$^{\circ}$\\
Volume&	879.4(3) \AA$^{3}$\\
$Z$&	4\\
Density (calculated)&	2.093 Mg m$^{-3}$\\
Absorption coefficient&	2.794 mm$^{-1}$\\
$F(000)$&	552\\
Crystal size&	$0.50 \times 0.32 \times 0.21$~mm$^{3}$\\
Crystal color and habit&	purple block\\
Diffractometer	&Bruker/Siemens SMART APEX\\
Theta range for data collection&	2.78 to 25.25$^{\circ}$ \\
Index ranges&	$-8 \leq h \leq 8$, $0 \leq k \leq 17$, $0\leq l \leq 10$\\
Reflections collected	&1984\\
Independent reflections&	1984 [$R(\mathrm{int}) = 0.0000$]\\
Completeness to theta = 25.25$^{\circ}$&	100.0 \% \\
Absorption correction&	Semi-empirical from equivalents\\
Max.\ and min.\ transmission&	0.556 and 0.315\\
Solution method&	SHELXS-97 (Sheldrick, 1990)\\
Refinement method&	Full-matrix least-squares on $F^{2}$\\
Data / restraints / parameters&	1984 / 30 / 149\\
Goodness-of-fit on F2&	1.072\\
Final $R$ indices [$I>2\sigma(I)$]&	$R1 = 0.0313, wR2 = 0.0841$\\
$R$ indices (all data)&$	R1 = 0.0332, wR2 = 0.0851$\\
Largest diff.\ peak and hole&	0.641 and -0.526~e.\AA$^{-3}$\\
\hline
\end{tabular}
\caption{Crystal data and structure refinement for
  Cu(gly)(pyz)(ClO$_{4}$) ({\bf 1}).}
\end{table*}

\section{X-ray structure determination of compound 2}
A suitable crystal was selected and was measured as described above, at
a temperature $T=296(2)$~K. 
 Data were measured using omega scans of 0.5$^{\circ}$ per frame for 10 seconds, and a 
full sphere of data was collected. A total of 1755 frames were
collected with a final resolution of 0.83~\AA.
 Cell parameters were retrieved using APEX2 \cite{jlm6}  software and refined
 using SAINTPlus \cite{jlm7}  on all observed reflections. 
Data reduction and correction for Lp and decay were performed using
the SAINTPlus software.
 Absorption corrections were applied using SADABS \cite{jlm8}.  The structure was
 solved by direct methods 
and refined by least squares method on $F^{2}$ using the SHELXTL program
package \cite{jlm9}.  The structure was 
solved in the space group $C2/c$ (no.\ 15) by analysis of systematic
absences. All non-hydrogen atoms were 
refined anisotropically. No decomposition was observed during data
collection. Details of the data 
collection and refinement are given in Table~2. Further x-ray measurements were made down to 100~K and no structural phase
changes were detected.

\begin{table*}
\begin{tabular}{cc}
\hline
Empirical formula &	C$_{8}$H$_{16}$Cl$_{2}$CuN$_{4}$O$_{13}$ \\
Formula weight &	510.69\\
Temperature &	296(2) K\\
Wavelength &	0.71073 \AA\\
Crystal system &	Monoclinic\\
Space group &	C2/c\\
Unit cell dimensions&	$a = 24.7388(16) $~\AA\\
&	$b = 6.8788(4)$~\AA\\	
&	$c = 10.2714(7)$~\AA\\	
&	$\alpha$= 90$^{\circ}$\\
&$\beta$= 95.669(4)$^{\circ}$\\
&$\gamma$ = 90$^{\circ}$\\
Volume&	1739.37(19)~\AA$^{3}$\\
$Z$&	4\\
Density (calculated)&	1.950 Mg m$^{-3}$\\
Absorption coefficient&	1.642 mm$^{-1}$\\
$F(000)$ &	1036\\
Crystal size&	0.22 x 0.21 x 0.05 mm$^{3}$\\
Crystal color and habit&	blue rod\\
Diffractometer&	Bruker/Siemens SMART APEX\\
Theta range for data collection&	3.07 to 25.25$^{\circ}$.\\
Index ranges&	$-29 \leq h \leq 29$, $-8 \leq k \leq 8$, $-12 \leq l
\leq 11$\\
Reflections collected&	13840\\
Independent reflections&	1581 [$R(\mathrm{int}) = 0.0280$]\\
Completeness to theta = 25.25$^{\circ}$&	99.9 \%\\ 
Absorption correction&	Semi-empirical from equivalents\\
Max. and min. transmission&	0.9224 and 0.7140\\
Solution method&	Bruker, 2003; XS, SHELXTL  v.6.14\\
Refinement method	& Full-matrix least-squares on $F^{2}$\\
Data / restraints / parameters &	1581 / 0 / 131\\
Goodness-of-fit on $F^{2}$&	1.080\\
Final $R$ indices [$I>2\sigma(I)$]&	$R1$ = 0.0336, $wR2$ = 0.0895\\
R indices (all data)&	$R1$ = 0.0349, $wR2$ = 0.0912\\
Extinction coefficient&	0.0100(7)\\
Largest diff. peak and hole&	0.477 and -0.414 e.\AA$^{-3}$\\
\hline
\end{tabular}
\caption{Crystal data and structure refinement for
  [Cu(gly)$_{2}$(pyz)(H$_{2}$O)](ClO$_{4}$)$_{2}$ ({\bf 2}).}
\end{table*}

\section{Static magnetic susceptibility and heat capacity}

The static magnetic susceptibility of {\bf 2} was
measured by use of a commercial SQUID (superconducting quantum
interference device) magnetometer (Quantum Design). The
magnetization of the sample was measured as a function of
temperature from room temperature down to 1.8~K at a constant
magnetic field of 5~mT.

The heat capacity was measured by use a thermal
relaxation method utilizing a commercial physical properties
measurement system (PPMS by Quantum Design) in a $^3$He cryostat.
The sample was glued to the heat-capacity platform by a small
amount of Apiezon N grease. The thermometer was carefully calibrated
for fields up to 14~T.

\section{Dynamic magnetic Susceptibility}

The dynamic magnetic susceptibility technique employs a radio
frequency (RF) circuit, 
a proximity detector oscillator (PDO) and, like most RF techniques, is based on an LCR circuit \cite{altarawneh}. 
The setup involves placing a single crystal sample in a small sensor
coil which is inductively coupled to the PDO using a coaxial cable. 
 The coil is placed on a measurement probe with a double axis rotator
 and inserted into the cryostat/magnet. 
To remove high frequency noise the output from the PDO chip is
amplified and put 
through a two-stage mixing and filtering process. 
A more detailed account of the setup can be found in Ref.~\cite{ghannadzadeh}.

Changes in the magnetization of the sample under an applied field
will lead to a change in the coil 
inductance which will modify the resonant frequency $\omega$ of the tank
circuit. For an insulating sample 
such as this, where the RF field penetrates the whole of the sample, the change in resonant frequency is given by
\begin{equation}
\Delta \omega = -a' \Delta \chi - b \Delta R_{0},
\end{equation}
with $a'=a f L_{\mathrm{empty}}$,
where $f$ is the coil filling factor, $L_{\mathrm{empty}}$ 
is the empty coil inductance, $\Delta \chi$ is the change in magnetic
susceptibility 
and $\Delta R_{0}$ is the magnetoresistance of the coaxial cable and
sensor coil in the presence of a magnetic field. 
The parameters $a$ and $b$ are positive constants.
To isolate the susceptibility the background contribution
($\Delta \omega_{\mathrm{bg}} = -b \Delta R_{0}$)
is measured with an identical empty coil and subtracted from the
sample measurements. 
Setting 
$\Delta \chi = \chi(B)-\chi_{0}$
 with constant $\chi_{0}$ gives \cite{ghannadzadeh}
\begin{equation}
\chi(B)=\frac{1}{a'} \left(\Delta \omega_{\mathrm{bg}}-\Delta \omega_{\mathrm{sample}} \right)+\chi_{0}.
\end{equation}

\section{Muon-spin relaxation measurements}

In a muon-spin relaxation ($\mu^{+}$SR) measurement \cite{sjb}
spin-polarized positive muons are
stopped in a target sample. The positive muons are attracted to
areas of negative charge density and often stop at interstitial
positions. The observed property of the experiment is the
time evolution of the muon-spin polarization, the behavior
of which depends on the local magnetic field at the muon site.
Each muon decays with an average lifetime of 2.2~$\mu$s into
two neutrinos and a positron, the latter particle being emitted
preferentially along the instantaneous direction of the muon
spin. Recording the time dependence of the positron emission
directions therefore allows the determination of the spin
polarization of the ensemble of muons. In our experiments,
positrons are detected by detectors placed forward (F) and
backward (B) of the initial muon polarization direction.
Histograms $N_{\mathrm{F}}(t)$ and $N_{\mathrm{B}}(t)$ record the number of positrons
detected in the two detectors as a function of time following
the muon implantation. The quantity of interest is the decay
positron asymmetry function, defined as
\begin{equation}
A(t) = \frac{N_{\mathrm{F}}(t) -\alpha N_{\mathrm{B}}(t)}
{N_{\mathrm{F}}(t) + \alpha N_{\mathrm{B}}(t)},
\end{equation}
where $\alpha$ is an experimental calibration constant. The asymmetry
$A(t)$ is proportional to the spin polarization of the muon
ensemble.

We carried out zero-field (ZF) measurements on {\bf 1} and 
transverse-field (TF) $\mu^{+}$SR measurements on {\bf  2} using
the LTF instrument at the Swiss Muon Source (S$\mu$S), Paul Scherrer
Institut, Switzerland. 
In the TF measurements, a magnetic field
is applied perpendicular to the initial muon spin direction,
causing a precession of the muon spins in the sum of the applied
and internal field directed perpendicular to the muon-spin
orientation. ZF and longitudinal-field (LF) measurements, where the 
field is applied parallel to the initial muon-spin direction, were made on
{\bf 2} using the 
HiFi instrument at the ISIS Facility, Rutherford Appleton Laboratory,
UK. 

\subsection{Sample 1: zero field $\mu^{+}$SR}
A powder sample of {\bf 1}
was mounted on the cold finger of a dilution refrigerator on the LTF instrument. The data (see Fig.~2 in the main text) were described well by the following fitting function across the whole temperature range
\begin{equation}
A(t)=A_1 e^{-\sigma_1^2t^2}+A_2 e^{-\sigma_2^2t^2}+A_3 e^{-\lambda t}+A_{\parallel},
\label{eqn:1d_fit}
\end{equation}
where the first term captures very rapid depolarization at early times ($\sigma_1\sim20-100$~MHz) probably due to muonium formation or molecular radical states of the muon, the second term captures the relaxation due to disordered nuclear moments ($\sigma_2=0.48$~MHz), the third term captures the relaxation due to electronic moments, and the non-relaxing final term is a combination of a background contribution and the fraction of muons polarized parallel to the local magnetic field.

\subsection{Sample 2: transverse field $\mu^{+}$SR}
\begin{figure}
\begin{center}
\epsfig{file=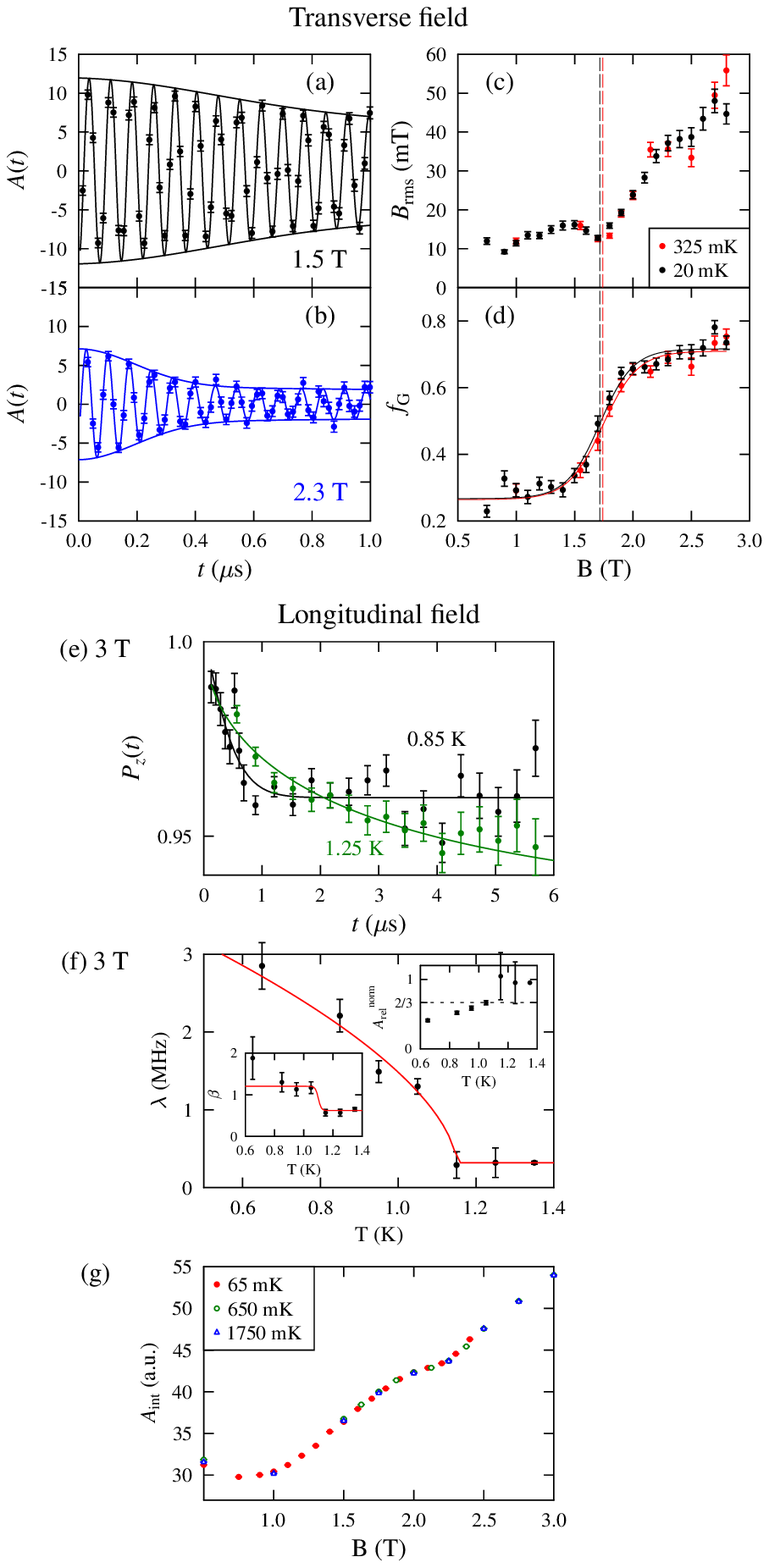,width=\columnwidth}
\caption{\label{fig:summary}(color online). (a)-(b) Transverse field data taken on LTF at 19~mK in (a) 1.5~T and (b) 2.3~T applied field. The data are shown in the `rotating reference frame` rotating at $B-0.1$~T, so oscillations are observed at $\gamma_\mu \times 0.1$~T~$=13.6$~MHz. (c) $B_{\rm rms}$ and (d) fraction of Gaussian relaxation $f_{\rm G}$ as a function of applied transverse field showing transitions at 1.72(6) and 1.74(8)~T at 20 and 325~mK, respectively. (e) Asymmetry in a longitudinal field of 3~T. (f) Relaxation rate against temperature at a LF of 3~T showing the transition at 1.15(5)~K, inset: temperature-dependence of beta (bottom) and the relaxing asymmetry (top). The relaxing asymmetry is normalised to its average for $T\ge1.15$~K. In (f) red lines are a guide to the eye. (g) Integrated asymmetry versus applied longitudinal field.}
\end{center}
\end{figure}

A polycrystalline sample was mounted on the cold finger of a dilution refrigerator at the LTF instrument. A magnetic field was applied at approximately 56$^\circ$ to the direction of the initial muon spin polarization and the transverse relaxation was followed as a function of applied field. The signals detected in two detectors placed on opposite sides of the sample were fitted simultaneously. In each of the detectors $d$, the positron count rate $N(d,t)$ was fitted to
\begin{align*}
\frac{N(d,t)-N_{\rm BG}(d)}{N_0 (d)\, e^{-t/\tau_\mu}}=& 1 + A(d) \cos [\gamma_\mu B\, t+\phi(d) ]  \\
& \times [ (1-A_{\rm G}) e^{-\lambda t} + A_{\rm G} e^{-(\gamma_\mu B_{\rm rms} t)^2/2} ],
\label{eqn:dimer_tf}
\end{align*}
where $N_{\rm BG}(d)$ is the background count rate, $N_0(d)$ the signal count rate, $A(d)$ is the asymmetry amplitude, $\phi(d)$ the detector phase, $\tau_{\mu}\approx2.2~\mu$s the muon lifetime, $\gamma_\mu$ the muon gyromagnetic ratio, $A_{\rm G}$ the fractional amplitude of the Gaussian relaxation, and 0.08~MHz~$< \lambda < 0.23$~MHz. We note that the transverse relaxation is described similarly well by two Gaussians (with different relaxation rates); this parameterization, however, provides a slightly better fit. Two signal fractions relaxing at different rates were, however, required to provide an accurate description of the data. Example asymmetry data are shown in Fig.~\ref{fig:summary}(a) and (b). The asymmetry $A(t)$ is given by 
\begin{equation*}
A(t)=\frac{N'(d_1,t)-\alpha N'(d_2,t)}{N'(d_1,t)+\alpha N'(d_2,t)},
\end{equation*}
where $N'(d,t)=N(d,t)-N_{\rm BG}(d)$ and $\alpha=N_0(d_1)/N_0(d_2)$ is a field-dependent experimental calibration constant accounting for different detector sensitivities. 

Fig.~\ref{fig:summary}(c) and (d) show $B_{\rm rms}$ and the Gaussian signal fraction $f_{\rm G}$ as a function of applied field $B$. A sharp increase in the relaxation and hence the distribution of (static) fields $B_{\rm rms}$ experienced by the muon is observed at 1.72(6) and 1.74(8)~T at 20 and 325~mK, respectively, concomitant with a sharp increase in the Gaussian-shaped relaxing fraction of the signal. The critical fields were obtained by fitting a finite-temperature step function to $f_{\rm G}$. This provides strong evidence of a transition to a long-range ordered state in the bulk of the sample. The observed signal is similar to that observed for field-induced transitions in the related material [Cu(HF$_2$)(pyz)$_2$]BF$_4$~\cite{steele} as well as the BEC candidate material Pb$_2$V$_3$O$_9$~\cite{carlo}. Note that a fraction of the signal ($\sim 25\%$) continues to relax at a much smaller rate in the ordered phase. This is most likely a background contribution due to muons stopped in the sample holder or cryostat tail.

\subsection{Sample 2: longitudinal field $\mu^{+}$SR}
In addition to the TF measurements performed on LTF,  LF data were
measured using the HiFi instrument \cite{lord_hifi}. Example asymmetry data are
shown in Fig.~\ref{fig:summary}(e). The longitudinal field was found
to rapidly quench any relaxation confirming the picture that the ZF
\musr\ relaxation is due to disordered static nuclear magnetism with
rapidly-fluctuating electronic moments that are motionally narrowed
from the spectra. Several field scans were performed to study the
changes in relaxation rate as well as in the time-integrated
asymmetry. 
Scans made as a function of applied field at fixed temperature were difficult to interpret, owing to a field-dependent
effect which causes dips in the integrated asymmetry 
to coincide with the minima of the beam spot size~\cite{lord_hifi}. 
As a result there is no resolvable temperature dependence over the range
65~mK--1.75~K [see Fig.~\ref{fig:summary}(g)].
However, it was possible to unambiguously identify the magnetic transition using a temperature scan performed at fixed longitudinal field of 3~T [see Fig.~\ref{fig:summary}(e) and (f)]. The asymmetry $A(t)$ was fitted to
\begin{equation}
A(t)=A_{\rm rel} \exp{[-(\lambda t)^\beta]} + A_{\rm BG},
\label{eqn:glydimer_lf}
\end{equation}
where the sum of relaxing amplitude $A_{\rm rel}$ and background
asymmetry $A_{\rm BG}$ was held constant. At a longitudinal field of
3~T, a sharp rise in the relaxation rate with a shape resembling that
expected for an order parameter was observed below
1.15(5)~K. Simultaneously the lineshape parameterised by $\beta$
changes, indicating a more static distribution of fields experienced
by the muon and the baseline asymmetry rises by approximately
$A_{\rm rel}(T\ge1.15~{\rm K})/3$ as expected for the development of
static magnetic order in a randomly-oriented polycrystalline
sample. The further drop of the relaxing asymmetry below 1.15~K is due
to the reduced instrument response to faster relaxation at a pulsed
muon source. Similar parameterisations of the data yield identical
results; the lineshape is well-described by a Gaussian for $T<1.15$~K
and a Lorentzian for $T\ge 1.15$~K. This demonstrates 
the sensitivity of LF \musr\ for the study of  a field-induced magnetic transition of this kind.

\section{Electron paramagnetic resonance}

\begin{figure}
\begin{center}
\epsfig{file=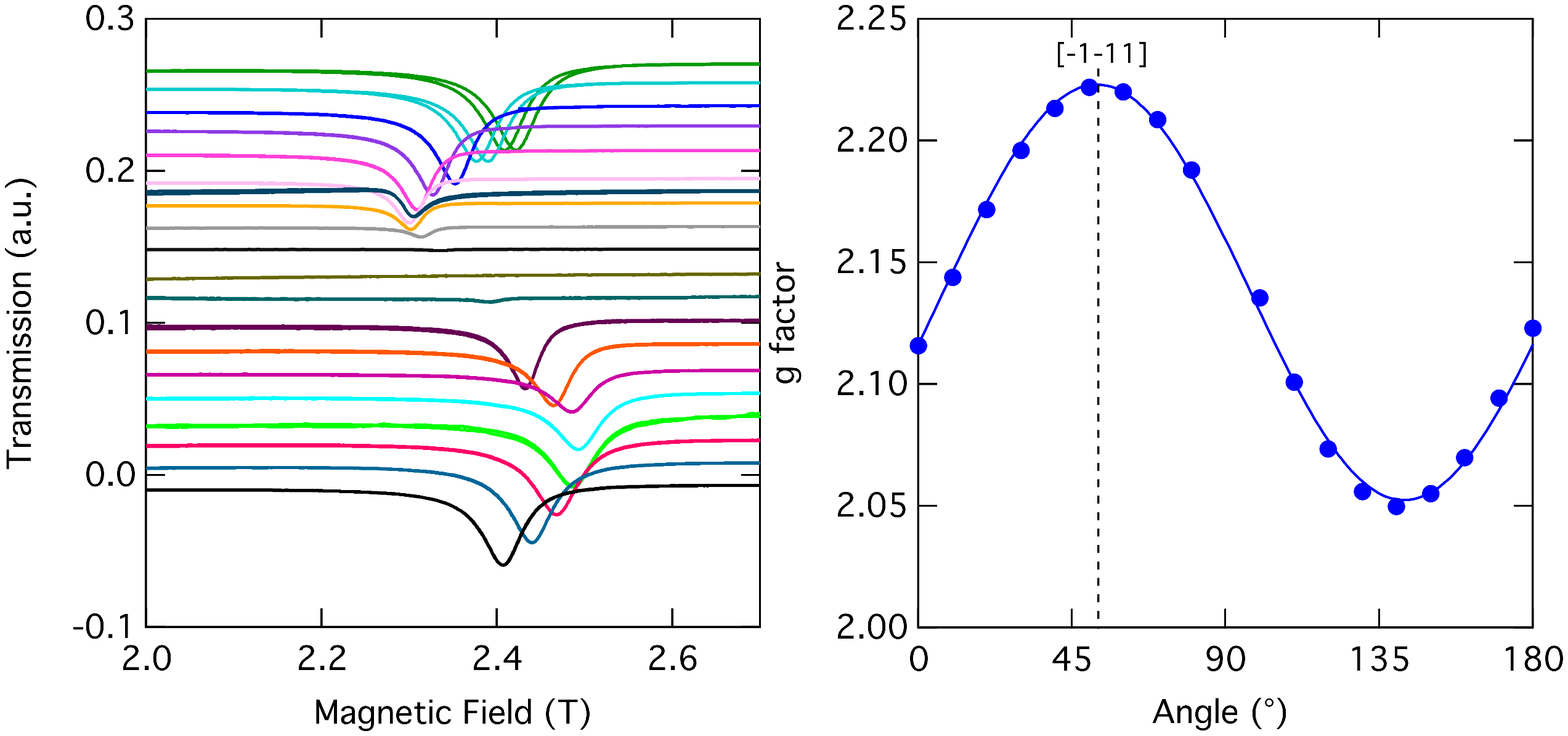, width=\columnwidth}
\caption{
a) EPR spectra of sample 2 measured at a temperature of 4~K and a
frequency of 70~GHz. The spectra correspond to discrete angles,  in
$10^{\circ}$
 increments offset for clarity, rotation through the principle magnetic axes. 
b) angular dependance of the $g$-factor measured at a temperature of 4~K for rotation through the principal magnetic axes.\label{eprfig}} 
\end{center}
\end{figure}
The electron paramagnetic resonance (EPR) spectra [Fig.~\ref{eprfig}(a)] were measured using
a cavity perturbation technique and an MVNA spectrometer manufactured
by AB-mm.  The angle dependance was performed using a mono-moded
cavity resonating at 70~GHz, that can be rotated with respect to the
applied magnetic field at cryogenic temperatures. Magnetic fields were
provided by a superconducting solenoid (up to 17~T). Standard $^4$He
techniques were employed to regulate temperature between 100~K and
1.5~K. No significant change in the $g$-factor or anisotropy was
detected in this temperature range. 
Multiple sets of rotations through high symmetry directions were used
to identify the approximately uniaxial 
nature of the $g$-tensor and its principle axes  [Fig.~\ref{eprfig}(b)].

\acknowledgments
The Bruker (Siemens) SMART APEX diffraction facility was established
at the University of Idaho with the assistance of the NSF-EPSCoR
program and 
the M. J. Murdock Charitable Trust, Vancouver, WA, USA.